\documentclass[prb,preprint,amssymb,amsfonts]{revtex4}
\usepackage{epsfig}
\usepackage{amsmath}

\begin{document}

\title{Dynamics of in-plane charge separation front in 2D electron-hole gas}

\author{Gang Chen, Ronen Rapaport, Steven H. Simon, Loren Pfeiffer, and Ken West}
\affiliation{Bell Laboratories, Lucent Technologies, 600 Mountain
Avenue, Murray Hill, New Jersey 07974}

\begin{abstract}
We show both experimentally and theoretically that the recently
observed optically induced in-plane charge separation in quantum
well (QW) structures and the exciton ring emission pattern at this
charge separation boundary have an extremely long lifetime. The
oppositely charged carriers remain separated and provide a
reservoir of excitons at their boundary with a persistent emission
which lasts hundreds of microseconds (orders of magnitude longer
than their recombination time) after the external excitation is
removed. This long lifetime is due to an interplay between the
slow in-plane carrier diffusion and slow carrier tunnelling
perpendicular to the QW plane.
\end{abstract}
\maketitle

Triggered by the quest for excitonic Bose-Einstein condensation
(BEC), recent studies on the emission from both double and single
GaAs based quantum well (QW) structures under tightly focused
optical excitation revealed a surprising ring pattern which
extends to a large distance away from the excitation spot
\cite{SnokeNature2002,ButovNature2002}. It was later found that an
optically induced in-plane separation of oppositely charged
plasmas is responsible for this excitonic ring emission pattern
formation \cite{RapaportPRL2004,ButovPRL2004}. The emerging
physical picture is as follows. Since the structures are n-doped
and in most experiments biased, there is an equilibrium density of
2D electron gas in the entire quantum well even in the absence of
optical excitation. The external optical excitation creates hot
electrons and holes above the barrier energy, some of which drift
across the QW to the contacts driven by the electrical field
applied perpendicular to the QW plane. The rest of the hot
electrons and holes cool down into the QW. However, this cooling
process is more likely for the holes due to their smaller drift
velocity (by a factor of 10-20 \cite{adachibook}) and shorter
phonon scattering time \cite{shahbook}. This leads to a depletion
of electrons that originally dwell in the QW near the laser
excitation spot via fast electron-hole recombination. While the
tunnelling of electrons into the QW tries to bring the electron
density back to the equilibrium level, high enough optical
excitation intensity will completely deplete the electrons and
eventually build up an excess population of holes at the
excitation spot, resulting in a hole puddle surrounded by a sea of
electrons outside the excitation spot in the plane of the QW
(perpendicular to the growth direction). Diffusion combined with
the Coulomb repulsion \cite{SnokePSS2004} of like charges drives
the holes in the puddle, and therefore the electron depletion
region, outwards. In the steady state, the densities of both
electron and hole plasmas at their boundary are low enough to
allow for the formation of excitons. The recombination of these
excitons gives rise to the large diameter ring pattern. It is
important to note that this is a different phenomenon from the
related giant ambipolar drift of spatially separated electrons and
holes in n-i-p-i superlattice structures \cite{Gulden1991,
Beck2001} (see endnote \cite{ambipolar} for more explanation).

The above picture portrays an interesting interplay of various
physical processes, such as carrier tunnelling (across the QW),
diffusion, drift (in the QW plane), and optical recombination.
These processes have distinct time scales and their interplay
directly determines the dynamics of the in-plane charge separation
and ring emission pattern. In this paper, we explore this dynamics
by measuring the response of the charge separation front to abrupt
changes in optical excitation intensity at the center spot. We
found that the overall reaction of the charge separation is
surprisingly slow: it takes the ring emission pattern hundreds of
microseconds to expand outward to the steady-state position after
turning on the optical excitation and hundreds of microseconds to
collapse into the center spot after turning off the excitation.
Theoretical analysis show that this extremely long lifetime is a
result of the slow carrier diffusion in the QW and tunnelling
across the QW. This extremely persistent charge separation also
provides an exciton reservoir that lasts orders of magnitude
longer than the exciton lifetime. In addition, since the optically
generated excess holes spend such a long time to diffuse to the
ring, they may have been well thermalized to the lattice
temperature before the formation of excitons.

As in our previous work \cite{RapaportPRL2004}, the sample
consists of a single 60\AA\ In$_{0.13}$Ga$_{0.87}$As quantum well
surrounded by GaAs/Al$_{0.32}$Ga$_{0.68}$As 50\AA/1000\AA\
barriers. A 1000 \AA\ layer of Si doped GaAs is located 2000 \AA\
from the QW on the $n^+$ substrate
 side and another similar layer is located 1000\AA\ from the QW on the top contact side.
Gold films are deposited on both sides of the sample to form
contacts. A 1 mm hole is opened on the top gold film for the
optical measurements. The sample was measured at 6K, and excited
with a HeNe laser (632 $nm$) with a spot diameter of $\sim 60\mu
m$.

\begin{figure}[h!]
\begin{center}
\includegraphics[scale=0.6]{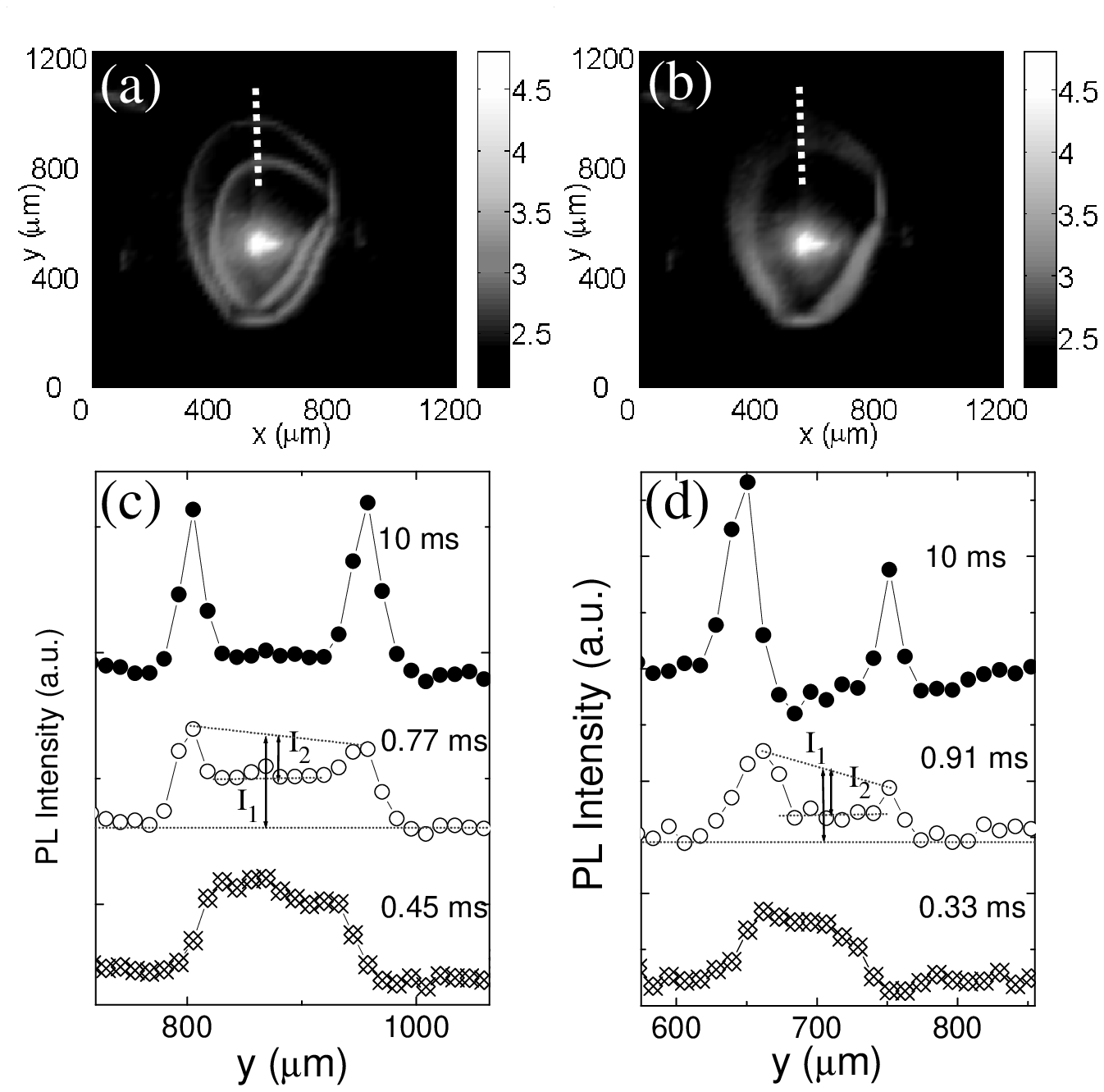}
\caption{PL images for light intensity modulation experiments at a
modulation period of  (a) 10 and (b) 0.33 ms (0.1 and 3 KHz). (c)
Emission profile along the dashed line in (a) and (b) for various
modulation periods, showing the transition from two concentric
rings to an annulus. (d) is similar to (c), except that the bias
voltage instead of the light intensity is modulated. The
schematics in the middle curves of (c) and (d) show how the ring
contrast, $C=\frac{I_2}{I_1}$, is defined.} \label{figure1}
\end{center}
\end{figure}

\begin{figure}[h!]
\begin{center}
\includegraphics[scale=0.6]{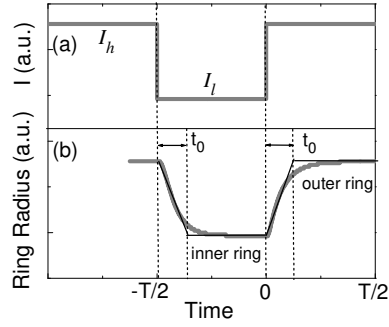}
\caption{(a) A schematic for the square modulation of the optical
excitation intensity between two values $I_h$ and $I_l$. (b) The
thick gray and thin dark lines describe the numerically calculated
and simplified charge separation boundary (ring) response within
one modulation cycle. It spends $t_0$ moving inward, another $t_0$
moving outward, and the rest of the time at the two steady-state
positions.}\label{figure2}
\end{center}
\end{figure}

In the particular experiments to be discussed below, the
excitation power ($I$) at the center was modulated, with a
modulation period $T$, between two values ($I_h=900\mu W$ and
$I_l=800\mu W$, square modulation), and with a fixed applied bias
voltage across the sample. The ring emission was imaged onto a
liquid nitrogen cooled CCD camera. Each image was collected over
an exposure time much longer than $T$. Figure.~\ref{figure1}
(a)-(b) show examples of collected images.
Figure.~\ref{figure2}(a) shows the square light intensity
modulation profile. At a modulation period of $T$=10 ms
(Fig.~\ref{figure1}(a)), two sharp concentric rings are clearly
observed with no emission in between. As $T$ is decreased (faster
modulation), the emission intensity in the region \textit{between}
the two original rings grows until the two concentric rings
transform into a wide emission annulus, as can be seen in
Fig.~\ref{figure1}(b) for $T$=0.33ms. This can be simply explained
as follows. For a very slow modulation, the transit time $t_0$ for
the charge separation front (observed as the ring emission) to
move from one steady state position to the other corresponding to
the two excitation intensities $I_h$ and $I_l$ is much smaller
than $T/2$. Therefore, one expects that the emission ring will
spend most of the time in the two steady state positions. Since
the image is integrated over many modulation cycles, most of the
light is expected to be emitted from two concentric rings. As $T$
decreases, the charge separation front spends relatively more time
in transit, leading to an increased emission from the region in
between the two steady state positions and therefore a decreased
contrast of the rings relative to the plateau in between. When
$t_0 \approx T/2$, the charge separation front spends all of its
time in transit ($t_0$ going outward and another $t_0$ going
inward), which diminishes the contrast of the rings and gives rise
to an annulus emission pattern. Experimentally, such transition
occurs around $T\sim 300-500 \mu s$ (a modulation frequency of
3-2KHz, depending on the applied bias), implying a surprisingly
long transit time $t_0$, on the order of hundreds of microseconds.

In a different experiment, the excitation intensity was fixed
while the applied bias was modulated between two values (similar
square wave modulation). The emission images are similar to those
taken in the light intensity modulation experiments. Again, a
transition from two concentric rings to a flat annulus at a
modulation period of hundreds of microseconds was observed. While
the excitation intensity modulation changes the carrier density
locally around the center excitation spot, the bias voltage
modulation changes several parameters across the entire sample,
including the carrier tunnelling time and the branching ratio of
the optically generated hot electrons and holes above the barriers
cooling into the QW. Therefore, the analysis of the light
modulation experiments are more straightforward.

Examples of the emission profiles showing the transition from two
concentric rings to a flat annulus are plotted in
Fig.~\ref{figure1}(c) and Fig.~\ref{figure1}(d) for the excitation
intensity and bias voltage modulation experiments, respectively.
We further define a measurable quantity, the contrast of the
concentric ring relative to the annulus, as $C=\frac{I_2}{I_1}$,
where $I_1$ is the \textit{average} peak emission intensity of the
two rings, and $I_2$ is the difference between $I_1$ and the
emission intensity of the plateau \textit{midway} between the
rings, as illustrated in the middle curves of
Fig.~\ref{figure1}(c)-\ref{figure1}(d). $C$ clearly depends on
both the modulation frequency and the ring response time $t_0$.
Based on previous discussions, $C$ should approach unit (maximum
contrast) at low modulation frequency and decrease as the
modulation frequency increase.

To show how $C$ should look in a simple way, we first assume that
the charge boundary moves at a constant speed between the two
steady state positions (we will show later why this assumption is
justified), spending a time of $t_0$ going outward and an
identical amount of time $t_0$ going inward in each modulation
cycle. This is shown by the thin dark solid line in
Fig.~\ref{figure2}(b). On average, the charge boundary spends
$2t_0$ in each modulation cycle in transit and the rest of the
time at the two steady state positions. For such a response
integrated much longer than $T$, the contrast $C$ should take the
form of $C=\frac{T/2-t_0}{T/2-t_0+2At_0}$ for $T\geq 2t_0$, where
$A$ is the ratio of the ring spatial width and the width of the
annulus. As expected, the contrast $C$ varies between 1 (two
concentric rings) and 0 (a flat annulus) as $T$ goes from infinity
to $2t_0$. For $T<2t_0$, the contrast should stay zero (flat
annulus).

\begin{figure}[h!]
\begin{center}
\includegraphics[scale=0.65]{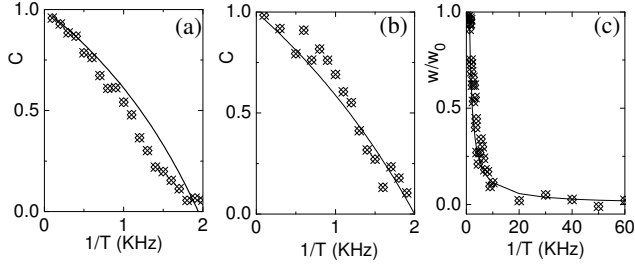}
\caption{Contrast $C$ of the two emission rings relative to the
annulus as a function of the modulation frequency for (a) the
light modulation and (b) the bias voltage modulation experiments.
The solid lines are fit using $C=(T/2-t_0)/(T/2-t_0+At_0)$, where
the ring response time $t_0$ is the only fitting parameter (see
the text). (c)Normalized annulus width $w$, as a function of
modulation frequency for the bias voltage modulation experiments.
The solid line represents $w/w_0=T/2t_0$, where $t_0$ is taken
from the fitting in (b).} \label{figure3}
\end{center}
\end{figure}

The measured contrast $C$ for the light and bias voltage
modulation experiments is plotted in Fig.~\ref{figure3}(a) and
Fig.~\ref{figure3}(b), respectively. The solid line in
Fig.~\ref{figure3}(a) shows the fitted $C$ for the light
modulation experiments with $A=0.25$ (obtained by measuring the
width of the annulus and the rings). The fitting yields a $t_0$ of
260 $\mu s$. For the voltage modulation experiments, $A$ is
$\sim$0.4 and the fitted solid line in Fig.~\ref{figure3}(b) gives
a $t_0$ of $250\mu s$.

For $T<2t_0$, the emission pattern remains a flat annulus.
However, since the charge separation front does not have enough
time to move the full range between the two steady state positions
for fast modulation, the annulus should narrow with decreasing
$T$. As an example, the annulus width, $w$, as a function of the
modulation frequency for the same set of bias modulation
experiments in Fig.~\ref{figure3}{b} is plotted in
Fig.~\ref{figure3}{c}. A simple calculation, again assuming that
the ring moves in and out at a constant speed, shows that the
dependence of $w$ on $T$ is of the form $w=w_0 \frac{T}{2t_0}$,
where $w_0$ is the radial distance between the two rings at low
modulation frequencies. We use the value for $t_0$ obtained from
Fig.~\ref{figure3}(b) and plot $w$ as a solid line in
Fig.~\ref{figure3}(c). It fits the data points very well.

We now discuss how this extremely long response time comes about
based on the model presented earlier and in Ref.
\cite{RapaportPRL2004,ButovPRL2004}. Generally speaking, the
charge separation boundary responds to the change of excitation
intensity on a time scale determined by both the carrier diffusion
and drift time in the QW plane and the tunnelling time through the
QW. Details of this dependence will be discussed later. In
Fig.~\ref{figure2}(b), the thick gray line shows the numerically
calculated ring emission response (using Eq.~\ref{n}-\ref{pcold}
described below) for a carrier tunnelling time of $50 \mu s$ and
an electron and hole diffusion coefficient of 20 $\mu m^2/ns$ and
2 $\mu m^2/ns$, respectively. The excitation light is modulated
between 800 and 900 $\mu W$ with a periodicity of T=400 $\mu s$.
The calculation clearly shows that the ring response time is
indeed extremely long. In addition, the ring does not expand and
contract at a constant speed: it moves faster right after the
abrupt change of excitation intensity and then gradually slow
down. However, at a particular ring position, the expansion and
contraction speeds roughly compensate such that the average speed
for every ring position during each modulation cycle almost does
not change. This justifies the earlier assumption that for images
integrated much longer than $T$ the charge separation boundary
looks as if it moves at a constant speed between the two
steady-state positions.

It is important to note that the above experiments were also
performed under different excitation intensity modulation depth.
We found that the ring response time does not vary much even for
modulations in which $I_l$ is almost zero and the charge boundary
moves all the way between the excitation spot and a large diameter
ring. These observations are consistent with the numerical
calculations.

We now discuss how $t_0$ depends on the carrier diffusion and
tunnelling times by analyzing the response of the cold electron
and hole densities ($n$ and $p$) in the QW determined by two
coupled rate equations
\cite{RapaportPRL2004,ButovPRL2004,SnokePSS2004} (neglecting the
Coulomb repulsion of like charges, which does not greatly alter
the numerical results qualitatively \cite{SnokePSS2004}):
\begin {eqnarray}
\frac{\partial n}{\partial
t}&=&D_{n}\nabla^2n-\frac{n-n_{eq}}{\tau_e} -\xi \,
n \,\, p+f_n \label{n}\\
\frac{\partial p}{\partial t}&=&D_{p}\nabla^2p-\frac{p}{\tau_h}
-\xi \, n\,\, p+f_q \label{pcold}
\end {eqnarray}Here, $D_e$ and $D_h$ are the electron and hole in-plane
diffusion-drift coefficients; $\tau_e$ and $\tau_h$ are the
electron and hole leakage (tunnelling) times; $f_n$ and $f_p$ are
the cold electron and hole sources due to the cooling of optically
generated hot carriers into the quantum well; $n_{eq}$ is the
constant equilibrium density of electrons in the QW in the absence
of the optical excitation; and $\xi$ is the electron-hole capture
coefficient. To simplify further analysis, we assume that
$D_e=D_h\equiv D$ and $\tau_n=\tau_h\equiv\tau$. The charge
imbalance $z=p-n$ then satisfies
\begin {equation}
\frac{\partial z}{\partial t}=D\nabla^2z -\frac{z-z_0}{\tau}+f_z
\label{nsimple}
\end {equation}where the original charge imbalance $z_0=-n_{eq}$ and $f_z=f_p-f_n$ is the charge imbalance source.
It is important to note that the exciton ring appears at the
contour of $z=0$ (the boundary between the electrons and holes).

\begin{figure}[h!]
\begin{center}
\includegraphics[scale=0.48]{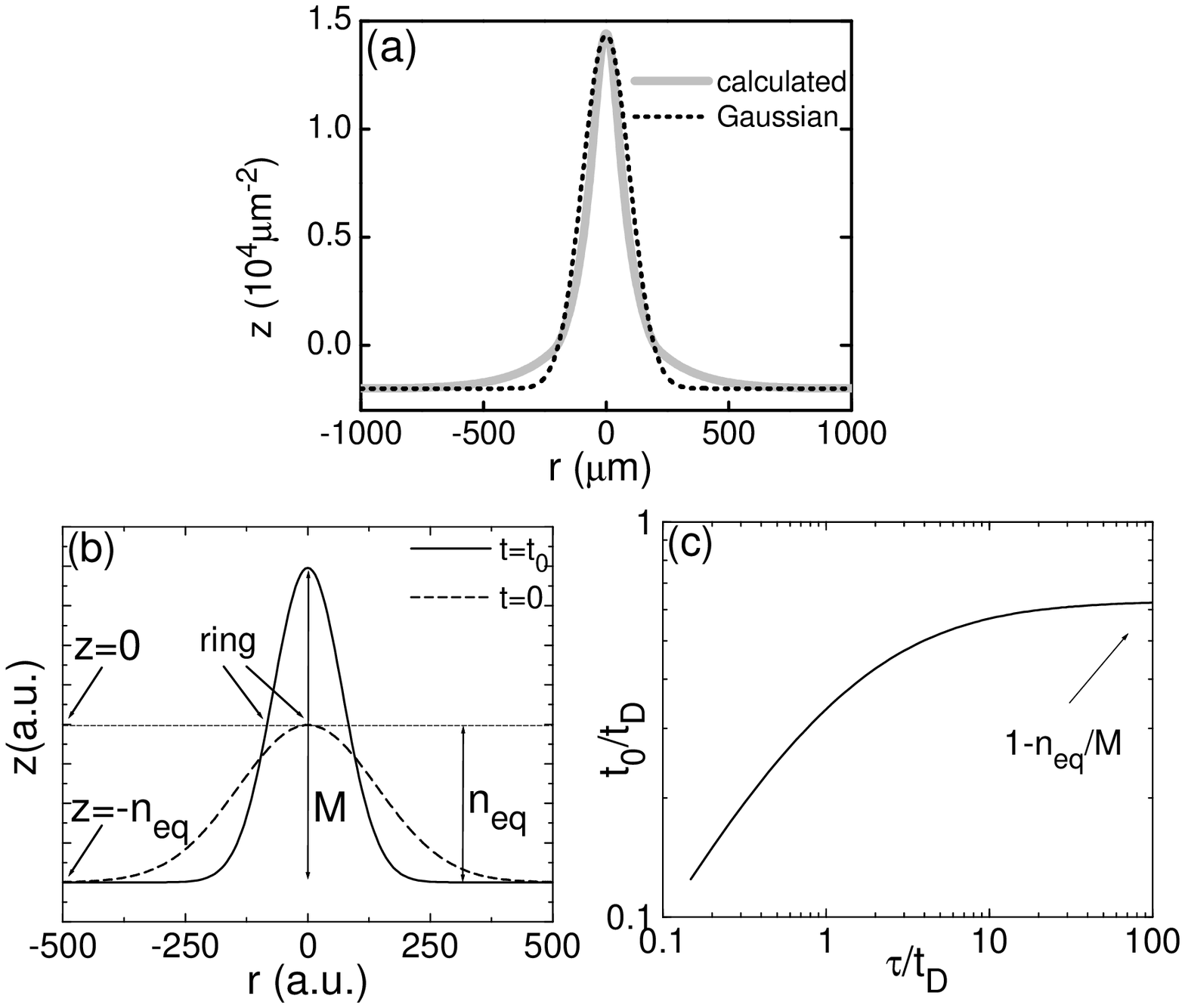}
\caption{(a) The solid gray line is the numerically calculated
steady-state charge imbalance density profile, $z$, under a
time-independent optical excitation (at r=0) which has a gaussian
profile with a full width half maximum of 60 $\mu m$. The dashed
line is a Gaussian with a peak value and an integrated charge
imbalance above $n_{eq}$ identical to the numerical calculation.
(b) Profiles of $z$ at $t=0$ and $t=t_0$. The optical excitation
at the center spot is turned off at $t=0$. The ring appears at
where $z=0$. The ring radius shrinks with time and at $t_0$ it
completely collapses. $M-n_{eq}$ is the peak $z$ at $r=0$ before
the optical excitation is turned off and $n_{eq}$ is the dark
equilibrium electron density in the QW. (c) $t_0$ as a function of
the tunnelling time $\tau$ in units of the characteristic
diffusion time $t_D$. $t_0$ is upper-bounded by $t_D(1-n_{eq}/M$)
for large $\tau$ and decreases with decreasing $\tau$.
\label{figure4}}
\end{center}
\end{figure}

The stationary solution of Eq.~\ref{nsimple} is shown by the gray
solid line in Fig.~\ref{figure4}(a) for a time \emph{independent}
excitation source at a small excitation spot with a Gaussian
profile (60 $\mu m$ full width half maximum). The charge imbalance
response dynamics can then be obtained by solving
Eq.~\ref{nsimple} using this stationary profile as the initial
condition and turning off the optical excitation ($f=0$). This
corresponds to our light modulation experiments with a large
modulation depth. If we now approximate this initial stationary
charge imbalance distribution as a Gaussian function
$z_{(r,t=0)}=Me^{-\frac{r^2}{\Delta^2}}-n_{eq}$ which has an
identical peak value and an integrated value above $n_{eq}$ (the
dashed Gaussian line in Fig.~\ref{figure4}(a); the validity of
this approximation will be discussed later), the solution of
Eq.~\ref{nsimple} becomes
\begin{equation}
z=n_{eq}\left(\frac{t_D}{t+\Delta^2/4D}e^{\frac{-r^2}{4Dt+\Delta^2}}e^{-\frac{t}{\tau}}-1\right),
\end{equation}where the diffusion
characteristic time is given by $t_D=M\Delta^2/4Dn_{eq}$.

Figure.~\ref{figure4}(b) shows this charge imbalance profile at
two different times. At t=0, it starts as a Gaussian and the
position of the emission ring is determined by $z=0$ (marked by
the arrows in Fig.~\ref{figure4}(b)). It then broadens and
decreases in amplitude. At $t=t_0$, where $t_0$ is the ring
response time discussed in the experiments, the charge imbalance
peak shrinks to just below $z=0$ and the ring collapses to the
center spot. $t_0$ satisfies $z(r=0,t=t_0)=0$, or
$t_0+t_Dn_{eq}/M=t_De^{-t_0/\tau}$, and is plotted in
Fig.~\ref{figure4}(c) as a function of $\tau$ in units of $t_D$.
We see that the ring response is determined by both the diffusion
time $t_D$ and the tunnelling time $\tau$. In particular, it shows
that $t_0$ is upper-bounded by $t_D(1-n_{eq}/M)$ for large $\tau$.
As $\tau$ becomes shorter, $t_0$ decreases and is eventually
determined by $\tau$ for $\tau<<t_D$. Note that $M$ is the
amplitude of the initial charge imbalance. It is obvious that
$M>n_{eq}$ (complete electron depletion at the excitation spot) is
a necessary condition for the ring formation and a positive $t_0$.

The experimental peak carrier imbalance density at the excitation
spot is roughly $M=2\times 10^{12}cm^{-2}$, which is extracted
from the measured linewidth of the emission
\cite{RapaportPRL2004}. The equilibrium carrier imbalance density
$z_0$ is extracted to be $10^{11}cm^{-2}$ by measuring the
linewidth of the emission at very low excitation intensity. The
characteristic diffusion time, $t_D$, is then calculated to be 250
$\mu s$ for a hole diffusion coefficient of $2 \mu m^2/ns$ and
$\Delta=300\mu m$ (estimated from an outer ring radius of 400 $\mu
m$. Assuming that the tunnelling time is much larger than $t_D$,
$t_0$ is then $\sim t_D(1-n_{eq}/M)=225\mu s$. This is in good
agreement with the experimentally measured value, particularly
considering various approximations we have made.

The above analytical solution of the ring response is based on the
assumption that the steady-state charge imbalance profile is a
Gaussian. The validity of this approximation is verified by the
identical $t_0$ numerically determined using either the
numerically simulated steady-state charge imbalanced profile or
its Gaussian approximation (Fig.\ref{figure4}(a)) as the initial
condition.

An important implication of the extremely long response time is
that after the excess holes are generated at the excitation spot,
they spend hundreds of microseconds migrating to where the
emission ring is. Therefore, it is expected that these initially
hot holes generated optically in the QW should have enough time to
thermalize to the lattice temperature. Compared to directly and
nonresonantly generated hot exctions,  the excitons formed at the
ring should have a kinetic energy that is no more than the exciton
binding energy. In addition, the supply of the excitons at the
ring is extremely persistent: the positive and negative charge
plasmas remain separated with excitons formed at their boundary
for microseconds even after the optical excitation source is
turned off.

We thank Phil Platzman, David Snoke, and Xing Wei for helpful
discussions.

\newpage
\vspace*{0cm}

\newpage
\vspace*{0cm}

\newpage
\vspace*{0cm}

\newpage
\vspace*{0cm}


\begin{thebibliography}{10}
\bibitem[{amb()}]{ambipolar}
\bibinfo{note}{In the n-i-p-i superlattice structures, the electrons and holes
  are separated along the growth direction driven by the built-in electric
  field. The enhanced diffusion coefficient and the negligible optical
  recombination (due to the electron-hole separation) are both critical to the
  giant carrier diffusion. In the QW structures, however, the electrons and
  holes are confined in the same QW and separated in-plane. A fast optical
  recombination is crucial for the electron depletion, the large charge
  separation, and the emission ring formation.}
\end{thebibliography}
\end{document}